\pdfoutput=1

\documentclass{svproc}
\usepackage{url}

\usepackage[utf8]{inputenc}
\usepackage{amssymb}
\usepackage{graphicx}
\usepackage{amsmath}
\usepackage{amssymb}
\renewcommand{\thefootnote}{$\star$}

\usepackage[pscoord]{eso-pic}
\usepackage{textcomp}
\usepackage{hyperref}
\newcommand{\placetextbox}[3]{
  \setbox0=\hbox{#3}
  \AddToShipoutPictureFG*{
    \put(\LenToUnit{#1\paperwidth},\LenToUnit{#2\paperheight}){\vtop{{\null}\makebox[0pt][c]{#3}}}%
  }%
}%

\begin{document}

\placetextbox{0.5}{0.96}{\normalfont \small Author accepted manuscript, published in ``Latent Variable Analysis and Signal Separation. LVA/ICA 2018. Lecture Notes in Computer Science,}

\placetextbox{0.5}{0.945}{\normalfont \small vol 10891, 568--577, 2018. Springer, Cham.}

\placetextbox{0.5}{0.930}{\normalfont \small The final authenticated publication is available online at \url{https://doi.org/10.1007/978-3-319-93764-9_52}.}

\mainmatter              

\title{Muticriteria decision making based on independent component analysis: A preliminary investigation considering the TOPSIS approach}
\titlerunning{Muticriteria decision making based on independent component analysis}  
%
\author{Guilherme Dean Pelegrina\inst{1}\thanks{The authors would like to thank the S\~{a}o Paulo Research Foundation (FAPESP - process n. 2016/21571-4 and 2017/23879-9) for the financial support. The authors also thank the National Council for Scientific and Technological Development (CNPq, Brazil) for funding their research.} \and Leonardo Tomazeli Duarte\inst{2} \and João Marcos Travassos Romano\inst{1}}
\authorrunning{G. D. Pelegrina, L. T. Duarte and J. M. T. Romano} 
%
\tocauthor{Guilherme Dean Pelegrina, Leonardo Tomazeli Duarte, Jo\~{a}o Marcos Travassos Romano}
\institute{School of Electrical and Computer Engineering (FEEC), University of Campinas (UNICAMP), Campinas, Brazil\\
\and
School of Applied Sciences (FCA), University of Campinas (UNICAMP), Limeira, Brazil \\
\email{pelegrina@decom.fee.unicamp.br}, \email{leonardo.duarte@fca.unicamp.br}, \email{romano@dmo.fee.unicamp.br}}

\maketitle              

\begin{abstract}

This work proposes the application of independent component analysis to the problem of ranking different alternatives by considering criteria that are not necessarily statistically independent. In this case, the observed data (the criteria values for all alternatives) can be modeled as mixtures of latent variables. Therefore, in the proposed approach, we perform ranking by means of the TOPSIS approach and based on the independent components extracted from the collected decision data. Numerical experiments attest the usefulness of the proposed approach, as they show that working with latent variables leads to better results compared to already existing methods.

\keywords{multi-criteria decision making, dependent criteria, independent component analysis, latent variables, TOPSIS}
\end{abstract}

\renewcommand{\thefootnote}{\arabic{footnote}}
\setcounter{footnote}{0}

\section{Introduction}
\label{sec:intro}

Many practical situations in multicriteria decision making (MCDM) consist in obtaining a ranking of a set of alternatives based on their evaluation according to a set of criteria~\cite{Figueira2005,Tzeng2011}. The main difference between the existing methods that perform ranking in MCDM is related to the criteria aggregation procedure. For instance, a natural way to perform aggregation is to consider a simple weighted sum~\cite{Tzeng2011} for all criteria and for a given alternative. Another strategy can be found in TOPSIS method (TOPSIS stands for Technique for Order Preferences by Similarity to an Ideal Solution)~\cite{Hwang1981}. In this method, one firstly defines a positive and a negative ideal alternative. Then, aggregation for a given alternative is done by calculating the Euclidean distances between the alternative under evaluation and the (positive and negative) ideal alternatives.

The original versions of the aforementioned approaches do not take into account any relation among criteria, which may lead to biased results in the aggregation step. Indeed, if, for instance, there are two criteria strongly correlated which are governed by a latent factor, then such a latent factor will have a strong influence on the aggregation step. In view of this inconvenient, there are some methods that try to deal with possible relations among the observed criteria~\cite{Grabisch1996,Antuche2010,Bondor2012,Vega2014,Wang2014}. Among them, an interesting approach is an extended version of TOPSIS~\cite{Antuche2010,Vega2014,Wang2014}. In this version, instead of considering the Euclidean distance in the aggregation step, one applies the Mahalanobis distance. Therefore, the calculation of the distance measure takes into account the covariance matrix among criteria.

However, a question that arises is whether the information about the covariance among criteria is sufficient to mitigate the biased effect of dependent criteria. Motivated by this question, this paper proposes a novel three-step procedure to deal with correlated criteria in decision making problems. In the first step of our proposal, we formulate the problem as a Blind Source Separation (BSS)~\cite{Comon2010} problem and apply an Independent Component Analysis (ICA) method to estimate the latent variables. The second step comprises the elimination of permutation and/or scale ambiguities provided by ICA. In the third step, we perform the TOPSIS approach based on the Euclidean distance on the estimated latent variables in order to obtain a global evaluation of the alternatives, thus allowing a final ranking. Aiming at verifying the proposed ICA-TOPSIS approach, we performed numerical experiments on synthetic data and compared the results obtained by our approach and the TOPSIS based on Mahalanobis distance. 

The rest of this paper is organized as follows. Section~\ref{sec:theory} discusses the main theoretical aspects about multicriteria decision making and blind source separation problems. Then, in Section~\ref{sec:proposal}, we present the proposed ICA-TOPSIS approach. The numerical experiments are described in Section~\ref{sec:exper}. Finally, in Section~\ref{sec:concl}, we present our conclusions and future perspectives.

\section{Theoretical background}
\label{sec:theory}

This section presents the theoretical aspects involved in multicriteria decision making and blind source separations problems.

\subsection{Multicriteria decision making problems and TOPSIS method}
\label{sec:mcdm}

The most relevant problems in MCDM consist in ranking a set of $K$ alternatives ($A = \left[ A_1, A_2, \ldots, A_K \right]$) based on a set of $M$ criteria ($C = \left[ C_1, C_2, \ldots, C_M \right]$). For each alternative $A_i$, $v_{i,j}$ represents its evaluation with respect to the criterion $C_j$. Therefore, in a MCDM problem, we often face with the following decision matrix (or decision data):

\begin{equation}
\label{eq:dec_data}
\mathbf{V}=\begin{array}{cc} 
 & \begin{array}{cccc} \, \, C_1 \, \, & \, \, C_2 \, & \ldots \, & \, C_M \, \, \\
\end{array} \\
\begin{array}{cccc}
A_1 \\
A_2 \\
\vdots \\
A_K
\end{array} & \left[\begin{array}{cccc}
v_{1,1} & v_{1,2} & \ldots & v_{1,M} \\
v_{2,1} & v_{2,2} & \ldots & v_{2,M} \\ 
\vdots & \vdots & \ddots & \vdots \\
v_{K,1} & v_{K,2} & \ldots & v_{K,M}
\end{array}
\right].
\end{array}
\end{equation}

Based on the decision matrix $\mathbf{V}$ and the set of weights $\mathbf{w}=[w_{1}, w_{2}, \ldots, w_{M}]$, which represent the ``importance'' of criterion $C_j$ in the decision problem, the goal is to aggregate $v_{i,j}$, $j=1, \ldots, M$ in order to obtain a global evaluation for each alternative $A_i$ and, then, to establish a ranking.

Several methods have been developed to deal with MCDM problems. Among them, a widely used one is the TOPSIS, developed by Hwang and Yoon~\cite{Hwang1981}. The main idea of this method is to determine the ranking based on the distances between each alternative and the (positive and negative) ideal solutions, as will be described in the sequel. The following steps describe the algorithm\footnote{We considered in this paper that all the criteria are to be maximized, i.e. the larger the better. However, if there are criteria to be minimized in the problem, some simple adaptations must be incorporated in the algorithm steps. For further details, please see~\cite{Hwang1981}.}:
\begin{enumerate}
\item The first step comprises the normalization of each evaluation $v_{i,j}$, given by
\begin{equation}
r_{i,j} = \frac{v_{i,j}}{\sqrt{\sum_{i=1}^K{v_{i,j}^2}}}, \, \, \, i=1, \ldots, K, \, j=1, \ldots, M.
\end{equation}
\item Based on $r_{i,j}$, we calculate the weighted normalized evaluation, given by
\begin{equation}
p_{i,j} = w_jr_{i,j}, \, \, \, i=1, \ldots, K, \, j=1, \ldots, M.
\end{equation}
\item In this step, we determine the positive ideal solution (PIS) and the negative ideal solution (NIS), given by
\begin{equation}
PIS = \mathbf{p}^+ = \left\{p_{1}^+, p_{2}^+, \ldots, p_{M}^+\right\},
\end{equation}
where $p_{j}^+=\max \{p_{i,j} | 1 \leq i \leq K \}$, $j=1, \ldots, M$, and
\begin{equation}
NIS = \mathbf{p}^- = \left\{p_{1}^-, p_{2}^-, \ldots, p_{M}^-\right\},
\end{equation}
where $p_{j}^-=\min \{p_{i,j} | 1 \leq i \leq K \}$, $j=1, \ldots, M$.
\item Given $PIS$ and $NIS$ derived in the last step, we calculate the distances (using Euclidean distance) from each evaluation vector $\mathbf{p}_i=[p_{i,1}, p_{i,2}, \ldots, p_{i,M}]$ representing alternative $A_i$ and both ideal solutions, described as follows:
\begin{equation}
D_i^+ = \sqrt{\left(\mathbf{p}_i-\mathbf{p}^+\right)^T\left(\mathbf{p}_i-\mathbf{p}^+\right)}, \, \, \, i=1, \ldots, K
\end{equation}
and
\begin{equation}
D_i^- = \sqrt{\left(\mathbf{p}_i-\mathbf{p}^-\right)^T\left(\mathbf{p}_i-\mathbf{p}^-\right)}, \, \, \, i=1, \ldots, K.
\end{equation}
\item In the last step, we determine the similarity measure of each alternative $A_i$ to the ideal solutions, given by
\begin{equation}
u_i = \frac{D_i^-}{D_i^+ + D_i^-}, \, \, \, i=1, \ldots, K,
\end{equation}
and derive the ranking according to $u_i$ in descending order.
\end{enumerate}

In this approach, one may note that the criteria are aggregated without taking into account any interaction between them. For example, in scenarios in which the criteria are correlated, i.e. they are composed by a combination of latent variables, disregarding the interaction may lead to biased results. In this context, an extended version of TOPSIS was proposed~\cite{Vega2014,Wang2014}, which takes into account the Mahalanobis distance~\cite{Mahalanobis1936} (instead of Euclidean distance) and, therefore, exploit the covariance among criteria. In this version, the distances calculated in step 4 are given by
\begin{equation}
\label{eq:dist_mah_pos}
DM_i^+ = \sqrt{\left(\mathbf{r}_i-\mathbf{r}^+\right)^T \mathbf{\Delta}^T\mathbf{\Sigma}^{-1} \mathbf{\Delta} \left(\mathbf{r}_i-\mathbf{r}^+\right)}, \, \, \, i=1, \ldots, K
\end{equation}
and
\begin{equation}
\label{eq:dist_mah_neg}
DM_i^- = \sqrt{\left(\mathbf{r}_i-\mathbf{r}^-\right)^T \mathbf{\Delta}^T \mathbf{\Sigma}^{-1} \mathbf{\Delta} \left(\mathbf{r}_i-\mathbf{r}^-\right)}, \, \, \, i=1, \ldots, K,\end{equation}
where $\mathbf{r}_i=[r_{i,1}, r_{i,2}, \ldots, r_{i,M}]$, $\mathbf{r}^+$ and $\mathbf{r}^-$ are, respectively, the positive and the negative ideal solutions derived from the normalized data $\mathbf{R}=\left(r_{i,j}\right)_{K \times M}$, $\mathbf{\Delta} = diag\left(w_1, w_2, \ldots, w_M\right)$ is the diagonal matrix whose elements are composed by the weights $\mathbf{w}$ and $\mathbf{\Sigma} \in \mathbb{R}^{M \times M}$ is the covariance matrix of $\mathbf{R}$. The similarity measure is calculated as described in step 5.

\subsection{Blind source separation problems and independent component analysis}

Let us suppose a set of signal sources $\mathbf{s}(k)=[s_1(k), s_2(k), \ldots, s_N(k)]$ that were linearly mixed according to
\begin{equation}
\label{eq:mix}
\mathbf{x}(k) = \mathbf{A}\mathbf{s}(k) + \mathbf{g}(k),
\end{equation}
where $\mathbf{A} \in \mathbb{R}^{M \times N}$ is the mixing matrix, $\mathbf{x}(k)=[x_1(k), x_2(k), \ldots, x_M(k)]$ is the set of mixed signals and $\mathbf{g}(k)=[g_1(k), g_2(k), \ldots, g_M(k)]$ is an additive white Gaussian noise (AWGN). In this linear case, BSS problems consist in retrieving the signal sources $\mathbf{s}(k)$ based only on the observed mixed data $\mathbf{x}(k)$, i.e. without the knowledge of both $\mathbf{s}(k)$ and mixing matrix $\mathbf{A}$~\cite{Comon2010}. This can be achieved by adjusting a separating matrix $\mathbf{B} \in \mathbb{R}^{N \times M}$ that provides a set of estimates $\mathbf{y}(k)=[y_1(k), y_2(k), \ldots, y_N(k)]$, given by
\begin{equation}
\label{eq:demix}
\mathbf{y}(k) = \mathbf{B}\mathbf{x}(k),
\end{equation}
which should be as close as possible from $\mathbf{s}(k)$. In this scenario, the separating matrix $\mathbf{B}$ should converge to the inverse of the unknown mixing matrix $\mathbf{A}$. However, given the permutation and scaling ambiguities inherent in BSS methods~\cite{Comon2010}, $\mathbf{B}$ may not be exactly the inverse of $\mathbf{A}$. As will be discussed latter on this paper, we made some assumptions on the problem in order to avoid these inconveniences.

There are several approaches used to deal with BSS problems. A common one, called ICA, is based on the assumption that the sources are i.i.d. (independent and identically distributed) and non-Gaussian. Given the mixing process expressed in~\eqref{eq:mix}, the observed sources are not independent anymore but close to Gaussian. Therefore, a simplified strategy to recover signal sources that are statistically independent is to formulate an optimization problem in which the cost function leads to the minimization of a Gaussian measure (e.g. kurtosis or negentropy) of the retrieved signals. An algorithm that is based on these assumptions is known as FastICA~\cite{Hyvarinen2001}. Another method that is used in BSS problems is the Infomax, proposed by Bell and Sejnowski~\cite{Bell1995}. This method, as demonstrated by Cardoso~\cite{Cardoso1997}, is closed-related to the maximum likelihood approach, which estimate the separating matrix $\mathbf{B}$ from the distribution of $\mathbf{x}(k)$. Both strategies will be used in our experiments.

\section{The proposed ICA-TOPSIS approach}
\label{sec:proposal}

In several problems in MCDM the criteria are dependent. For example, consider the case of determining a ranking of $K$ students evaluated according to their grades in sociology, mathematics and physics\footnote{It is worth mentioning that this MCDM problem is addressed by other works in the literature~\cite{Grabisch1996,Kojadinovic2008}}. It is possible that both grades in mathematics and physics are correlated criteria, since they usually measure similar competences. Therefore, the aggregation based on the collected data may lead to biased results. In this case, one may think that a proper analysis should be made in the latent variables $\mathbf{l}(k)=[l_1(k), l_2(k), \ldots, l_N(k)]^T$ associated with the collected data $\mathbf{V}$ through the mixing process
\begin{equation}
\label{eq:mix_prop}
\mathbf{V}^T = \mathbf{A}\mathbf{l}(k) + \mathbf{g}(k),
\end{equation}
where $\mathbf{A} \in \mathbb{R}^{M \times N}$ represents the mixing process acting on the latent variables $\mathbf{l}(k)$ and $\mathbf{g}(k)=[g_1(k), g_2(k), \ldots, g_M(k)]$ is an additive white Gaussian noise (AWGN). One may note that equation~\eqref{eq:mix_prop} is similar to~\eqref{eq:mix}, with $\mathbf{l}(k)$ and $\mathbf{V}^T$ representing, respectively, the set of signal sources and the mixed signals. Therefore, aiming at performing the MCDM analysis on the latent variables, as mentioned in Section~\ref{sec:intro}, the application of Mahalanobis distance in TOPSIS approach may not be sufficient to deal with dependent criteria, since only the information of covariance among criteria is taken into account. 

In this context, this paper proposes to deal with the problem of dependent criteria in MCDM applying an ICA-TOPSIS approach, which comprises three steps. In the first one, we formulate a BSS problem whose aim is to recover the latent variables based on the mixed decision data $\mathbf{V}$. In this formulation, we consider that the number of criteria is equal to the number of latent variables, which leads to the determined case $M=N$ in BSS. Therefore, after estimating the separating matrix $\mathbf{B}$, we obtain the estimated latent variables $\mathbf{\hat{l}}(k)=[\hat{l}_1(k), \hat{l}_2(k), \ldots, \hat{l}_N(k)]^T$, given by
\begin{equation}
\label{eq:demix_prop}
\mathbf{\hat{l}}(k) = \mathbf{B}\mathbf{V}^T,
\end{equation}
similarly as described in~\eqref{eq:demix}.

The second step comprises the adjustment of the estimated latent variables in order to avoid permutation and/or scale ambiguities. In this procedure, we made the assumption that the diagonal elements in the mixing matrix $\mathbf{A}$ is positive and greater, in absolute value, than all the off-diagonal elements in the same row, i.e. each latent variable has a positive majority influence in each mixed criterion. Therefore, based on the separating matrix $\mathbf{B}$ and, consequently, on the estimated mixing matrix $\mathbf{\hat{A}}=\mathbf{B}^{-1}$, we perform the following adjustment\footnote{It is worth mentioning that the scale ambiguity provided by a positive factor or a negative factor different from $-1$ is automatically mitigated in the normalization step of TOPSIS.}:

\begin{itemize}
\item For the first row in $\mathbf{\hat{A}}$, we find the column $q$ in which the greater absolute value is located. Therefore, we permute the first and the $q$ columns of $\mathbf{\hat{A}}$. In order to correctly resetting the estimated latent variables, we also permute the first and the $q$ estimates. After repeating this procedure for all rows in $\mathbf{\hat{A}}$, we obtain the estimated mixing matrix partially adjusted $\mathbf{\hat{A}}^{Adj_{p}}$ and avoid the permutation ambiguity provided by the BSS method.
\item Based on the assumption that the diagonal elements in the mixing matrix $\mathbf{A}$ is positive, if a diagonal element $q'$ of $\mathbf{\hat{A}}^{Adj_{p}}$ is negative, we multiply all the elements in the same column of $q'$ by $-1$. This leads to the signal inversion of the estimated latent variable $\hat{l}_{q'}$, since equation~\eqref{eq:mix_prop} needs to be valid. After verifying all the diagonal elements of $\mathbf{\hat{A}}^{Adj_{p}}$ and performing the signal changes, we obtain the final adjusted estimated mixing matrix $\mathbf{\hat{A}}^{Adj_{f}}$ and avoid the scale ambiguity provided by the $-1$ factor. 
\end{itemize}

In order to illustrated these adjustments, suppose that we achieve the estimated mixing matrix
$$\mathbf{\hat{A}}=\left[\begin{array}{cc}
1.52 & -2,95 \\
2.01 & 0.85 \\
\end{array} \right]$$
associated with the retrieved sources $\mathbf{\hat{l}}(k)=[\hat{l}_1(k), \hat{l}_2(k)]^T$. Based on our assumptions, the first adjustment leads to
$$\mathbf{\hat{A}}^{Adj_{p}}=\left[\begin{array}{cc}
-2.95 & 1.52 \\
0.85 & 2.01 \\
\end{array} \right],$$
and to the retrieved sources partially adjusted $\mathbf{\hat{l}}^{Adj_{p}}(k)=[\hat{l}_2(k), \hat{l}_1(k)]$. One may note the permutation of both columns. In the second adjustment, we obtain
$$\mathbf{\hat{A}}^{Adj_{f}}=\left[\begin{array}{cc}
2.95 & 1.52 \\
-0.85 & 2.01 \\
\end{array} \right]$$
and $\mathbf{\hat{l}}^{Adj_{f}}(k)=[-\hat{l}_2(k), \hat{l}_1(k)]$, which corrects the signal of the retrieved sources. 

After performing the ICA and eliminating the ambiguities, the third step of the proposed approach comprises the application of TOPSIS based on Euclidean distance in $\mathbf{\hat{l}}^{Adj_{f}}(k)$ and the ranking determination. 

\section{Numerical experiments}
\label{sec:exper}

Aiming at verifying the application of the proposed ICA-TOPSIS approach to deal with dependent criteria in MCDM problems, we performed numerical experiments based on synthetic data and compared the results with the ones provided by existing methods. The next section describes the considered data and the obtained results.

\subsection{Data generation}

In this paper, we performed the experiments based on a decision data comprised by 100 alternatives and 2 criteria, both with the same importance ($w_1=w_2=0.5$). The latent variables were randomly generated according to a uniform distribution in the range $[0,1]$. In order to derive the ``collected'' observed data $\mathbf{V}$, we considered the mixing matrix
$$\mathbf{A}=\left[\begin{array}{cc}
1.00 & -0.15 \\
0.30 & 1.00 \\
\end{array} \right]$$
and the mixing process described in~\eqref{eq:mix}, in which $\mathbf{s}(k)$ and $\mathbf{x}(k)$ represent the latent variables and the observed data $\mathbf{V}$, respectively. Moreover, the additive noise was applied considering a Signal-to-Noise Ratio (SNR), given by
\begin{equation}
SNR = 10 \log_{10} \frac{\sigma_{signal}^2}{\sigma_{noise}^2},
\end{equation}
where $\sigma_{signal}^2$ and $\sigma_{noise}^2$ are, respectively, the signal power and the noise power, in the range $(0,50]$.

\subsection{Comparison between the considered approaches}

In order to verify the application of the proposal, we first generate the latent variables and derive the ranking according to the original TOPSIS method (based on Euclidean distance). This ranking is considered as the correct one, since it is obtained directly from the (unknown) latent variables. Therefore, we perform the mixing process and, given the mixed observed data, we apply the proposed ICA-TOPSIS approach (based on FastICA and Infomax algorithms), the original TOPSIS and the TOPSIS based on Mahalanobis distance. The obtained results are compared according to a performance index called normalized Kendall tau distance~\cite{Kendall1938}, which calculates the percentage of pairwise disagreements between two rankings. This measure is defined by
\begin{equation}
\tau=\frac{N_D}{K(K-1)/2},
\end{equation}
where $N_D$ is the number of pairwise disagreements between the rankings and $K$ is the number of alternatives. Therefore, $\tau$ close to zero indicates that there is no disagreement between the two rankings, i.e. the obtained ranking is the same that the correct one provided by the original TOPSIS method applied on the latent variables.

\begin{figure}[ht]
\centering
\includegraphics[height=6.2cm]{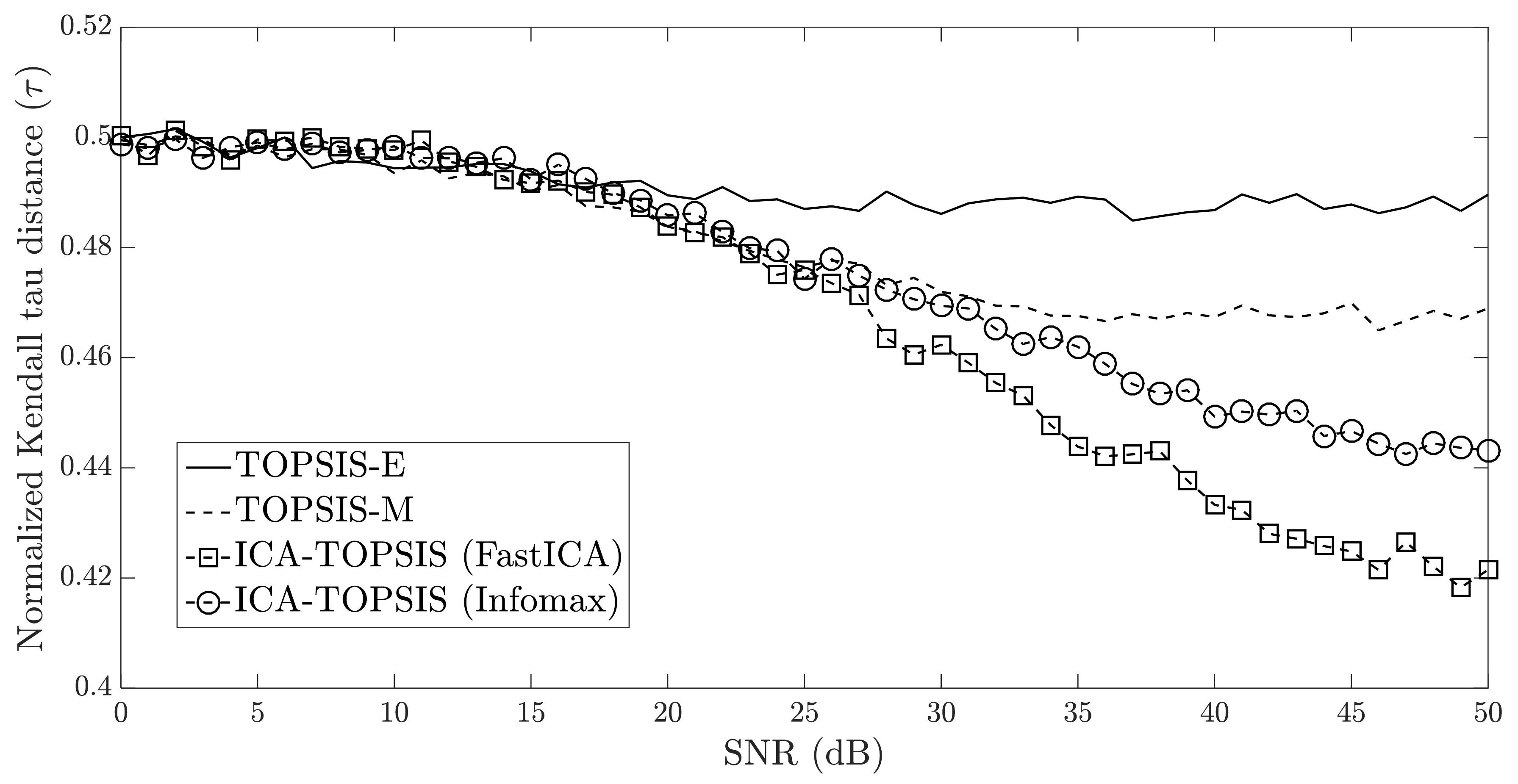}
\caption{Comparison of the Kendall tau distances for the original TOPSIS based on Euclidean distance (TOPSIS-E), the TOPSIS based on Mahalanobis distance (TOPSIS-M) and the proposed approach (with FastICA and Infomax).}
\label{fig:kendall}
\end{figure}

Figure~\ref{fig:kendall} presents the Kendall tau distance for each considered method and SNR value (averaged over 1000 realizations). One may note that the TOPSIS based on Mahalanobis distance improves the original version of this method, leading to lower values of $\tau$. However, the best results were obtained applying the ICA-TOPSIS, specially for SNR values greater than 25 dB. In terms of the FastICA and Infomax algorithms, the former achieved a better performance.

\section{Conclusions and perspectives}
\label{sec:concl}

Dependent criteria is an important issue in multicriteria decision making. In order to deal with this problem, several methods has been developed, such as the TOPSIS based on Mahalanobis distance. In this work, we presented preliminaries discussions on a novel approach used to mitigate biased results provided by dependent criteria. This approach, called ICA-TOPSIS, comprises the application of independent component analysis in order to extract the latent variables from the observed decision data and, then, the use of the original TOPSIS to derive the ranking based on the retrieved independent data.

Based on the MCDM scenario considered in this work and the obtained results, one may remark that the proposed ICA-TOPSIS approach leads to better results compared to the methods found in the literature. For instance, our proposal achieved lower Kendall tau values compared to the TOPSIS based on Mahalanobis distance, which is used in several works in the literature. A possible explanation for this result is that the ICA methods exploit the independence among criteria, which is stronger than the covariance information used in TOPSIS based on Mahalanobis distance. Since we consider a MCDM problem comprised by a mixture of latent variables, our proposal can better mitigate the biased effect of the criteria dependence. 

It is worth mentioning that this work presented initial results on the application of ICA-TOPSIS approach to deal with MCDM problems. Future works comprise a further understanding on this proposal, especially on the latent variable estimation step. Different numbers of criteria and alternatives will also be considered in new experiments. Moreover, we aim at verifying the performance of the proposed approach on decision problems based on real data.

%
%

\end{document}